\documentclass{PoS}

\title{Semileptonic decays of heavy-light pseudoscalar mesons}

\ShortTitle{Semileptonic decays of heavy-light pseudoscalar mesons}

\author{\speaker{Nazario Tantalo}\thanks{based on work done in collaboration with
        G.M. de Divitiis, E. Molinaro and R. Petronzio}%
        \\
        INFN "Tor Vergata" and E. Fermi Ctr., Rome\\
        E-mail: \email{nazario.tantalo@roma2.infn.it}}

\abstract{
I discuss the results of a recent quenched lattice calculation of
the two independent form factors parametrizing the semileptonic
decays between heavy-light pseudoscalar mesons.
The differential decay rate of the process $B\rightarrow D\ell\nu$
has been calculated at non vanishing momentum transfer both in the
case of the light leptons $\ell=e,\mu$ and in the case of a non
vanishing lepton mass $\ell=\tau$.
}

\FullConference{The XXV International Symposium on Lattice Field Theory\\
                July 30 - August 4 2007\\
                Regensburg, Germany}

\begin{document}

\section{Introduction}
In this talk I discuss the results of a recent quenched lattice calculation of the matrix elements
of the vector part of the heavy-heavy weak currents between pseudoscalar heavy-light meson
states (see refs.~\cite{de Divitiis:2007ui,de Divitiis:2007uk}). 
These matrix elements are parametrized in terms of two independent form factors whose accurate
knowledge is required in order to extract the matrix element $V_{cb}$ of 
the CKM~\cite{Cabibbo:1963yz,Kobayashi:1973fv} matrix from the experimental 
measurements of the differential decay rates of the semileptonic
decays $B\rightarrow D\ell\nu_{\ell}$. In the case of the light leptons, $\ell=e,\mu$,
the differential decay rate is proportional to the square of a particular linear combination 
of the two form factors, usually called $G^{B\rightarrow D}(w)$. 
Both the form factors are needed in the case of the heavy $\tau$ 
lepton~\cite{Kiers:1997zt,Chen:2006nua}. The BaBar and Belle collaborations 
have already measured~\cite{Aubert:2007ab,Matyja:2007kt} 
the branching ratios of the processes $B\rightarrow D^{(*)}\tau\nu_{\tau}$ and
a future measurement of the differential decay rate will make possible to extract
$V_{cb}$ also from this channel. Within the heavy quark effective theory (HQET) it has been 
shown~\cite{Isgur:1989ed} that the semileptonic transitions between heavy-light mesons
can be parametrized, at leading order of the expansion in the inverse heavy quark mass, 
in terms of a single universal form factor known as Isgur-Wise function.
The Isgur-Wise function is universal in the sense that it describes any semileptonic
decay mediated by heavy-heavy weak currents regardless of the flavour of the initial
and final heavy quarks and of the spins of the mesons. From the phenomenological point
of view it is relevant to know the size of the corrections to the Isgur-Wise limit and
to establish at which order the heavy quark expansion has to be truncated to
produce useful results down to the charm mass.

In refs.~\cite{de Divitiis:2007ui,de Divitiis:2007uk} the form factors have been
calculated at non vanishing momentum transfer for many different combinations of
the initial and final heavy quark masses ranging from the physical bottom quark
mass to the physical charm quark mass.
The simulation of relativistic heavy quarks has been performed by using  
the step scaling method (SSM)~\cite{Guagnelli:2002jd}, already applied 
successfully to the determination of heavy 
quark masses and heavy-light meson decay 
constants~\cite{deDivitiis:2003wy,deDivitiis:2003iy,Guazzini:2006bn}. 
The SSM allows to reconcile large quark masses with 
adequate lattice resolution and large physical volumes.
The two form factors have been calculated for different values of the momentum transfer by making 
use of flavour twisted boundary conditions~\cite{deDivitiis:2004kq}, 
that shift the discretized set of lattice momenta by an arbitrary amount 
(see also~\cite{Martinelli:1982bm,Bedaque:2004kc,Sachrajda:2004mi,Flynn:2005in}).

\section{Form factors}
Semileptonic decays of pseudoscalar mesons into pseudoscalar mesons 
are mediated by the vector part of the weak $V-A$ current and the corresponding matrix elements
can be parametrized in terms of two form factors,
\begin{eqnarray}
\frac{\langle \mathcal{M}_f\vert \ V^\mu \ \vert \mathcal{M}_i\rangle}{\sqrt{M_i M_f}}=
(v_i+v_f)^\mu \ h_+^{i\rightarrow f} + (v_i-v_f)^\mu \ h_-^{i\rightarrow f}
\label{eq:hphmdef}
\end{eqnarray}
where $v_{i,f}=p_{i,f}/{M_{i,f}}$ are the $4$-velocities of the mesons. 
The form factors depend upon the masses of the parent and daughter particles and upon
$w\equiv v_f\cdot v_i$
\begin{eqnarray}
h_\pm^{i\rightarrow f}(w)\equiv h_\pm(w,M_i,M_f),
\nonumber
\end{eqnarray}
Time reversal and hermiticity imply that $h_+^{i\rightarrow f}$ and
$h_-^{i\rightarrow f}$ are real. 
Furthermore they imply that $h_+^{i\rightarrow f}$ is even under the 
interchange of the initial and final states while $h_-^{i\rightarrow f}$ is odd,
\begin{eqnarray}
h_+(w,M_i,M_f)\, =\,h_+(w,M_f,M_i), \qquad \qquad
h_-(w,M_i,M_f)\, =\,-\ h_-(w,M_f,M_i)
\label{eq:evenodd}
\end{eqnarray}
As a consequence, the elastic form factor $h_-(w,M_i,M_i)$ vanishes identically.
Concerning the form factor $h_+(w,M_i,M_f)$, it has a well defined limit when
both $M_i$ and $M_f$ are sent to infinity at fixed ratio $r=M_f/M_i$.
It is thus legitimate to make a change of variables from the meson masses to the parameters 
$\varepsilon_+$ and $\varepsilon_-$, defined as
\begin{eqnarray}
\varepsilon_+=\frac{1}{M_f}+\frac{1}{M_i},
\qquad \qquad
\varepsilon_-=\frac{1}{M_f}-\frac{1}{M_i}
\end{eqnarray}
and expand $h_+(w,\varepsilon_+,\varepsilon_-)$ in Taylor series around the
point $\varepsilon_{\pm}=0$
\begin{eqnarray}
h_+(w,\varepsilon_+,\varepsilon_-) &=& h_+(w,0,0) + 
\varepsilon_+ \frac{\partial h_+(w,0,0)}{\partial \varepsilon_+} +
\frac{\varepsilon_+^2}{2} \frac{\partial^2 h_+(w,0,0)}{\partial \varepsilon_+^2} +
\frac{\varepsilon_-^2}{2} \frac{\partial^2 h_+(w,0,0)}{\partial \varepsilon_-^2} + \dots
\nonumber
\end{eqnarray}
The conservation of the vector current implies that $h_+^{i\rightarrow i}(w=1)=1$ and
the previous relation, at zero recoil, can be rewritten in the form
\begin{eqnarray}
h_+(w=1,\varepsilon_+,\varepsilon_-) &=& 1 + 
\frac{\varepsilon_-^2}{2} \frac{\partial^2 h_+(w=1,0,0)}{\partial \varepsilon_-^2} +
\dots
\label{eq:taylorexp1}
\end{eqnarray}

The semileptonic decay rate of a $B$ meson into a $D$ meson, in 
the approximation of massless leptons $\ell=e,\mu$, is given by
\begin{eqnarray}
&&\frac{d\Gamma^{B\rightarrow D\ell\nu_{\ell}}}{dw}=
\vert V_{cb}\vert^2 \frac{G_F^2}{48\pi^3}(M_{B}+M_{D})^2M_{D}^3(w^2-1)^{3/2}
\left[ G^{B\rightarrow D}(w)\right]^2,
\nonumber \\
&&1 \le w \le \frac{M_B^2+M_D^2}{2M_BM_D}
\end{eqnarray}
where the form factor $G^{B\rightarrow D}(w)$ is related to
$h_+^{i\rightarrow f}(w)$ and $h_-^{i\rightarrow f}(w)$ by
\begin{eqnarray}
G^{i\rightarrow f}(w)=h^{i\rightarrow f}_+(w)\ -\
\frac{M_f-M_i}{M_f+M_i}\ h^{i\rightarrow f}_-(w)
\nonumber
\end{eqnarray}
In the case $\ell=\tau$ the mass of the lepton cannot be neglected and the differential
decay rate is given by~\cite{Korner:1989qb,Kiers:1997zt}
\begin{eqnarray}
&&\frac{d\Gamma^{B\rightarrow D\tau\nu_{\tau}}}{dw}=
\frac{d\Gamma^{B\rightarrow D (e,\mu)\nu_{e,\mu}}}{dw}
\left(1-\frac{r_{\tau}^2}{t(w)} \right)^2
\left\{
\left(1+\frac{r_{\tau}^2}{2t(w)} \right)
+\frac{3r_{\tau}^2}{2t(w)}\frac{w+1}{w-1} \left[\Delta^{B\rightarrow D}(w)\right]^2
\right\}
\nonumber \\ 
&& r_{\tau}=\frac{m_{\tau}}{M_B}, \qquad r=\frac{M_D}{M_B}, \qquad t(w)=1+r^2-2rw,
\nonumber \\ 
&&1 \le w \le \frac{M_B^2+M_D^2-m_{\tau}^2}{2M_BM_D}
\nonumber 
\end{eqnarray}
where
\begin{eqnarray}
\Delta^{i\rightarrow f}(w) =
\frac{1}{G^{i\rightarrow f}(w)}
\left[\frac{1-r}{1+r}\ h^{i\rightarrow f}_+(w)\ -\ \frac{w-1}{w+1}\ h^{i\rightarrow f}_-(w) \right]
\label{eq:deltadef}
\end{eqnarray}
In the elastic case $\Delta^{i\rightarrow f}(w)$ vanishes identically and, in
the approximation in which $h^{i\rightarrow f}_-(w)$ is much smaller than $h^{i\rightarrow f}_+(w)$,
it is very well approximated by its static limit
\begin{eqnarray}
\Delta^{i\rightarrow f}(w) \simeq \frac{1-r}{1+r},\qquad\qquad r=\frac{M_f}{M_i}
\label{eq:deltastatic}
\end{eqnarray}
All the details on the lattice definitions of the form factors are given
in refs.~\cite{de Divitiis:2007ui,de Divitiis:2007uk}.
\begin{figure}[t]
\includegraphics[width=\textwidth]{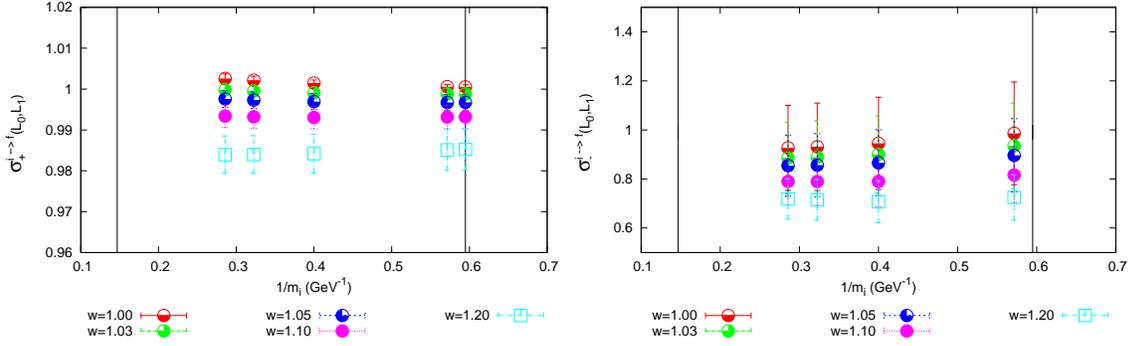}
\caption{\label{fig:ssf1} Step scaling functions  of $h_+^{i\rightarrow c}$ (left)
and $h_-^{i\rightarrow c}$ (right) as functions of $1/m_i$ for the first
evolution step (from $L_0$ to $L_1$). The black vertical
lines represent the physical points $m_i=m_c$ and $m_i=m_b$.
The data are in the continuum and chiral limits.} 
\end{figure}
\begin{figure}[t]
\includegraphics[width=\textwidth]{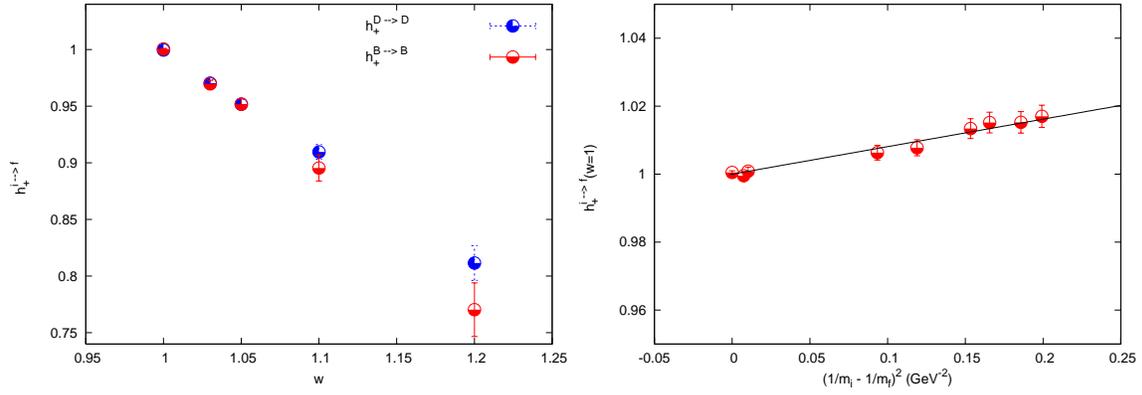}
\caption{\label{fig:iw2} 
The left plot shows $h_+^{B\rightarrow B}(w)$, $h_+^{B\rightarrow D}(w)$
and $h_+^{D\rightarrow D}(w)$: in the range $1\le w \le 1.05$ 
the two elastic form factors are indistinguishable within the quoted errors while 
$h_+^{B\rightarrow D}(w)$ shows appreciable corrections from the Isgur-Wise limit, in 
particular at zero recoil.
The right plot shows $h_+^{i\rightarrow f}$ at zero recoil ($w=1$) as a function
of $\varepsilon_-^2$ (actually $(1/m_i-1/m_f)^2 \propto \varepsilon_-^2$,
$m_{i,f}$ being the RGI heavy quark masses). } 
\end{figure}
%

\section{Step Scaling Method}
The SSM has been introduced to cope with two-scale problems in lattice QCD. In the
calculation of heavy-light meson properties the two scales are the mass
of the heavy quarks ($b$,$c$) and the mass of the light quarks ($u$,$d$,$s$). 
In describing how the SSM works in the present case, I consider the generic form factor 
$F^{i\rightarrow f}=\{h_+^{i\rightarrow f},h_-^{i\rightarrow f},G^{i\rightarrow f}\}$
as a function of $w$, the volume $L^3$ and the meson states.
The last are fixed by the corresponding
heavy and light RGI quark masses that, being extracted by the lattice version of the 
PCAC relation, are not affected by finite volume effects. 
The first step of the finite volume recursion consists in calculating
the observable $F^{i\rightarrow f}(w;L_0)$ on a small volume, $L_0=0.4$~fm, 
which is chosen to accommodate the dynamics of heavy quarks with masses ranging from the 
physical value of the charm mass up to the mass of the bottom.
A first effect of finite volume is taken into account by evolving
the results from $L_0$ to $L_1=0.8$~fm through the factor
\begin{displaymath} 
\sigma^{i\rightarrow f}(w;L_0,L_1)=\frac{F^{i\rightarrow f}(w;L_1)}{F^{i\rightarrow f}(w;L_0)}
\end{displaymath}
computed for each value of $w$ and for each value of the light quark mass.
The crucial point is that the step scaling functions are calculated by simulating
heavy quark masses smaller than the $b$-quark mass; more precisely,
the step scaling functions at $m_i\simeq m_b$ and $m_f\simeq m_c$
are obtained by directly simulating $m_f$ both on $L_0$ and on $L_1$ and
by a smooth extrapolation in $1/m_i$. 
Extrapolating the step scaling functions is more advantageous than extrapolating 
the form factors. This can be easily understood by relying on HQET expectations 
(see also eq.~(\ref{eq:taylorexp1})),
\begin{eqnarray} 
\sigma^{i\rightarrow f}(w;L_0,L_1) &=&
\frac{F^{(0) \rightarrow f}(w;L_1)}{F^{(0) \rightarrow f}(w;L_0)}
\;\left[
1+\frac{F^{(1) \rightarrow f}(w;L_1)-F^{(1) \rightarrow f}(w;L_0)}{m_i}
+\dots\right]
\end{eqnarray}
In the previous relations the superscripts in parenthesis, $(n)$, 
mark the order of the expansion in the inverse heavy quark mass.
The subleading correction to the step scaling functions is the difference
of two terms and vanishes in the infinite volume, 
becoming smaller and smaller as the volume is increased.
This matches the general idea that finite volume effects, measured
by the $\sigma$'s, are almost insensitive to the high energy scale.
In order to remove the residual finite volume effects the procedure
described above is iterated once more, passing from $L_1$ to $L_2=1.2$~fm.
Final results are obtained from
\begin{eqnarray} 
F^{i\rightarrow f}(w;L_2) \quad=\quad
F^{i\rightarrow f}(w;L_0)\quad
\sigma^{i\rightarrow f}(w;L_0,L_1)\quad
\sigma^{i\rightarrow f}(w;L_1,L_2)
\label{eq:ssm}
\end{eqnarray}
\begin{figure}[t]
\begin{center}
\includegraphics[width=0.8\textwidth]{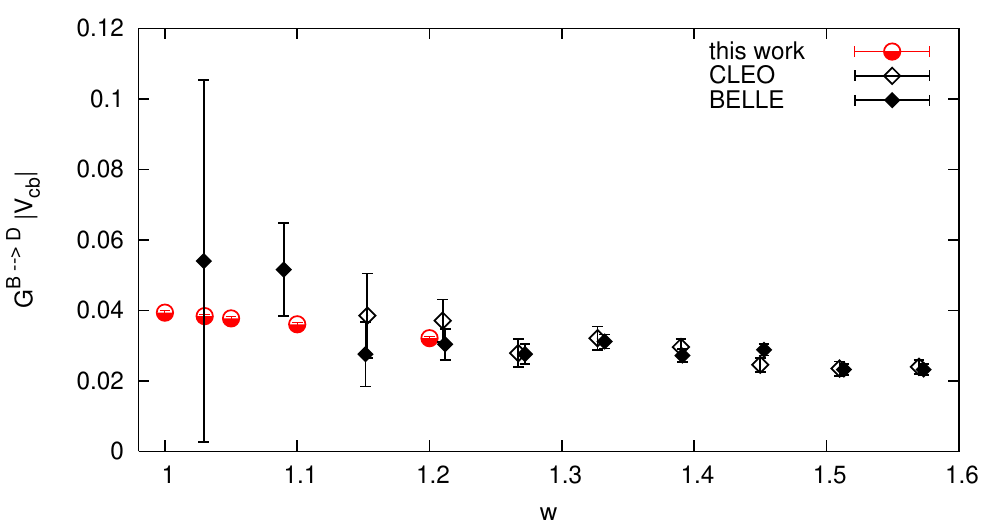}\\
\includegraphics[width=0.8\textwidth]{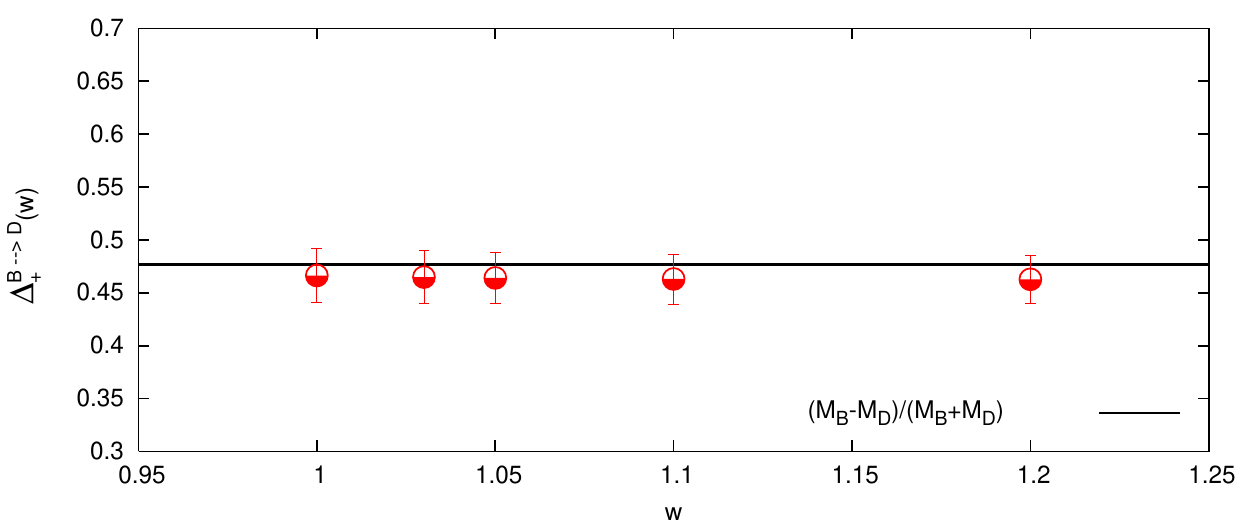}
\end{center}
\caption{\label{fig:vcb} Upper plot: comparison of $|V_{cb}|\ G^{B\rightarrow D}(w)$ 
with available experimental data; the plot has been done by normalizing lattice data 
with the value of $V_{cb}$ extracted at $w=1.2$.
Lower plot: the figure shows the function $\Delta^{B\rightarrow D}(w)$
in the chiral, continuum, and infinite volume limits; the solid line correspond to the static limit
result, $(M_B-M_D)/(M_B+M_D)$, and has been drawn by using the experimental determinations
of the meson masses.} 
\end{figure}
%

\section{Results}
In figure~\ref{fig:ssf1} one can test the hypothesis on
the low sensitivity of the step scaling functions upon the high energy scale.
The figure shows the step scaling functions of the form factors $h_+^{i\rightarrow c}$ (left)
and $h_-^{i\rightarrow c}$ (right) as functions of $1/m_i$. In both cases the dependence
upon $m_i$ is hardly appreciable and in the case of $h_+^{i\rightarrow c}$ the $\sigma$'s are 
very close to one while $h_-^{i\rightarrow c}$ is affected by stronger finite volume effects. 
The values at $m_i=m_b$ are obtained by linear fits. Similar plots for the other form
factors and for the second evolution step can be found in 
refs.~\cite{de Divitiis:2007ui,de Divitiis:2007uk}.

Eq.~(\ref{eq:taylorexp1}) predicts that the convergence toward
the static limit is faster in the case of the elastic form factors with respect to the
ones having $m_i>m_f$. This happens because near the point at zero recoil the subleading
corrections to $h_+^{i\rightarrow f}(w)$ are proportional to the square of the difference
of the initial and final meson masses.
Figure~\ref{fig:iw2} clearly shows that this happens in practice. Indeed, in the left plot,
the elastic form factor $h_+^{D\rightarrow D}(w)$ is much closer to
the static limit (very well approximated by $h_+^{B\rightarrow B}(w)$) with respect to
the form factor $h_+^{B\rightarrow D}(w)$, 
the one relevant into the calculation of $V_{cb}$.
In the right plot of figure~\ref{fig:iw2} one can see how well 
eq.~(\ref{eq:taylorexp1}) is approximated by numerical data. The fit is performed on the
slope while the intercept is fixed to one. 

Concerning $G^{B\rightarrow D}(w)$, figure~\ref{fig:vcb} (upper plot)
shows our results for values of $w$ up to $w\simeq 1.2$ and physical 
$b$ and $c$ quark masses. 
The comparison with available measurements has been done by extracting the value of $V_{cb}$ by 
the ratio of the experimental and lattice data at $w=1.2$; as an indication, we get 
$V_{cb}=3.84(9)(42)\times10^{-2}$,
where the first error is from the lattice result, $G^{B\rightarrow D}(w=1.2)=0.853(21)$,
and the second from the experimental decay rate.
In the lower plot of figure~\ref{fig:vcb} I show our best result for the function $\Delta^{B\rightarrow D}(w)$
that enters in the decay rate of the process $B\rightarrow D\tau\nu_{\tau}$.
$\Delta^{B\rightarrow D}(w)$ does not show any significant dependence upon
$w$ and is very well approximated by its static limit (see eq.~\ref{eq:deltastatic}).
These findings represent a prediction that can be confirmed by a future measurement of the 
differential decay rate of the process $B\rightarrow D\tau\nu_{\tau}$. 
Indeed, the function $\Delta^{B\rightarrow D}(w)$
can be extracted experimentally by the ratio 
$d\Gamma^{B\rightarrow D\tau\nu_{\tau}}/d\Gamma^{B\rightarrow D (e,\mu)\nu_{e,\mu}}$ that does not
depend upon the CKM matrix element. On the other hand, the knowledge of $\Delta^{B\rightarrow D}(w)$
is required in order to perform lepton-flavour universality checks on the extraction
of $V_{cb}$.

\acknowledgments{The results discussed in this talk have been
obtained in collaboration with G.M. de Divitiis, E. Molinaro and R. Petronzio
to whom goes my warm thanks.
An email exchange with M. Della Morte and R. Sommer
on the subjects covered in this talk it is gratefully acknowledged.}



\end{document}